\documentstyle[prb,twocolumn, aps]{revtex}
\input epsf
\begin{document}

\smallskip 
\twocolumn[\hsize\textwidth\columnwidth\hsize\csname@twocolumnfalse\endcsname

\title{Transient current spectroscopy of a quantum dot in the Coulomb blockade regime}
\author{Toshimasa Fujisawa$^{1}$, Yasuhiro Tokura$^{1}$, and Yoshiro Hirayama$^{1,2}$}
\address{$^{1}$NTT Basic Research Laboratories, 3-1 Morinosato-Wakamiya, Atsugi, 243-0198, Japan\\
$^{2}$CREST, 4-1-8 Honmachi, Kawaguchi, 331-0012, Japan}
\date{Received }
\maketitle

\begin{abstract}
Transient current spectroscopy is proposed and demonstrated in order to
investigate the energy relaxation inside a quantum dot in the Coulomb blockade regime. 
We employ a fast pulse signal to excite an AlGaAs/GaAs
quantum dot to an excited state, and analyze the non-equilibrium transient
current as a function of the pulse length. The amplitude and time-constant of the
transient current are sensitive to the ground and excited spin states. 
We find that the spin relaxation time is longer than, at least, a few $\mu $s.
\end{abstract}


] \narrowtext

\newpage 

Electron spin in a semiconductor quantum dot (QD) has been intensively
studied as a candidate for quantum information storage in the solid state
and single spintronics in which a single spin state determines the transport
through the dot.\cite{Loss,Awschalom,Gupta} Electronic QDs, in which
discrete energy states are filled up with a controllable number of
electrons, show various spin-related phenomena, including spin degeneracy,
exchange interaction, spin blockade, and Kondo physics. \cite
{QDotReview,TaruchaAA,LPKexcitationAA,Sasaki} These conventional dc
transport studies measure the overlap of the dot state to the reservoir.
However, the transitions between these dot states, which are responsible for
the coupling to the environment, should strongly reflect the characteristics
of the states by analogy with the selection rules for absorption and
emission processes in real atoms. Electronic states in a solid are easily
coupled to the lattice environment, while atomic states couple strongly to
an electromagnetic environment. Theory predicts that the energy relaxation
time of an excited state (ES) in a QD is of the order of nanoseconds if the
transition is accompanied by acoustic phonon emission, provided the spin is
neglected.\cite{Bockelmann} By contrast, it should be extremely long if the
transition involves a spin flip.\cite{Khaetskii} Spin conservation during a
transition is a selection rule for artificial atoms. Several experimental
reports address the relaxation in QDs, however, the relaxation is considered
to be an electron-phonon interaction without spin-flip.\cite
{JWeis,Agam,FujisawaScience}

Here we report on a pulse-excited transport measurement designed to estimate
the energy relaxation time from an ES to the ground state (GS) in a QD. We
find that the relaxation time is short (less than a few ns) if the
transition conserves the total spin, while it is very long (more than a few $%
\mu $s) if the transition involves a spin-flip. We also find that the spin
state can be investigated using transient current spectroscopy because of
the spin conservation rule.

A QD (addition energy $E_{add}$ of about 2 meV, single particle energy
spacing $\Delta $ of 100 - 300 $\mu $eV, number of electrons $N<$ 50) is
weakly coupled to the left and the right leads [see Fig. 1(a)].\cite
{FujisawaScience,Oosterkamp,RFSET} We made the tunneling rate of the left
barrier, $\Gamma _{L}$, much larger than that of the right barrier, $\Gamma
_{R}$, by roughly tuning the gate voltages, $V_{L}$ and $V_{R}$. Fine tuning
of $V_{R}$ effectively changes the electrostatic potential of the dot
allowing us to see one of the Coulomb blockade (CB) oscillations, as shown
in Fig. 2(a). If the incoming tunnel rate is smaller than the outgoing
tunnel rate (negative bias voltage, $V_{b}<0$, in our experiments), the
current, $I$, is given by the rate $\Gamma _{R,i}$ through the thick barrier
to the $i$-th empty state located in the transport window (between the left
and right Fermi energies, $\mu _{L}$ and $\mu _{R}$, respectively); $I=e\sum
\Gamma _{R,i}$. The current increases stepwise when an energy level, $E_{i}$%
, enters the transport window at $\mu _{R}=E_{i}$, as indicated by the solid
lines. All the energy states in a QD can be detected in this way, unless the
coupling to the leads is too weak. We can also estimate each rate $\Gamma
_{R,i}$ from the appropriate current step heights.\cite{QDotReview}

The magnetic field ($B$) dependence of the excitation spectrum $-dI/dV_{R}$
at $V_{b}$ = -1 mV (small incoming rate) is shown in Fig. 3(a). The GS and
ESs appear as positive peaks, which are plotted separately in Fig. 3(d). The
orbital characteristics of the states are manifested by the $B$ dependence
on energy and the current amplitude (``magnetic fingerprint'').\cite
{TaruchaAA,LPKexcitationAA,Stewart} The complicated $B$ dependence is a
result of crossings or anti-crossings between different energy states in the
QD. The first ES in the $N$-electron stripe has the same $B$ dependence as
the GS of the ($N+1$)-electron stripe. Similar correlations in neighboring
stripes between $N-1$ and $N+2$, and no even-odd effects due to spin
degeneracy are seen in the spectrum, as reported for low-symmetry QDs.\cite
{Stewart} Even though the Zeeman energy (about 20 $\mu $eV/T) is much
smaller than $\Delta $, many-body effects give rise to the appearance of
spin-polarized states even in a weak magnetic field. The energy relaxation
from an ES to the GS strongly depends on their spin states.

In order to investigate the energy relaxation time from such ESs to the GS,
we use a pulse excitation to push the system out of equilibrium. We apply a
pulse signal to the right gate to modulate the electrostatic potential of
the dot [see Figs. 1(a)-1(c)], and measure the non-equilibrium transient
current in the following way. During the steady state before the pulse [Fig.
1(d)], both the ES and the GS for a particular number of electrons are
located above $\mu _{L}$ and $\mu _{R}$ for a sufficiently long period to
ensure that these states are completely empty. The positive pulse lowers the
potential of the dot by 100 - 800 $\mu $eV, which is smaller than $E_{add}$.
This means the GS and the ES cannot be simultaneously occupied due to CB. If
the potential is lowered so that only the ES is located in the transport
window [Fig. 1(e)], transport through the ES continues until the GS becomes
occupied. First, an electron tunnels into the ES or the GS with a
probability ratio, $\Gamma _{L,e}:\Gamma _{L,g}$. If an electron is injected
into the ES, it can relax to the GS, or tunnel to the right lead to give a
net current. The transport is transient, being blocked once the GS is
occupied. We chose asymmetric barriers, $\Gamma _{L}\gg \Gamma _{R}$, such
that an electron can stay in the ES for a long time. Thus, the decay time, $%
\tau $, of the transient current should reflect the energy relaxation rate, $%
W$, from the ES to the GS.

We can relate $\tau $\ and $W$ from the rate equations, 
\[
\frac{d}{dt}\rho _{e}=\Gamma _{L,e}(1-\rho _{e}-\rho _{g})-\Gamma _{R,e}\rho
_{e}-W\rho _{e}
\]
\begin{equation}
\frac{d}{dt}\rho _{g}=(\Gamma _{R,g}+\Gamma _{L,g})(1-\rho _{e}-\rho
_{g})+W\rho _{e},  \label{EqRateeq}
\end{equation}
by taking into account all the tunneling processes shown schematically in
Fig. 1(e). Here $\rho _{e}$ and $\rho _{g}$ are the average electron numbers
in the ES and the GS, respectively ($\rho _{e}=\rho _{g}=0$ at the beginning
of the pulse $t=0$, and $0\leq \rho _{e}+\rho _{g}\leq 1$ is satisfied due
to CB). We obtain a form $\rho _{e}(t)=A(1-$e$^{-Ft})$e$^{-Dt}$, assuming $%
\Gamma _{L}\gg \Gamma _{R}$ and $W<\Gamma _{L,e}+\Gamma _{L,g}$ (valid for
the visible current). $A\simeq \frac{\Gamma _{L,e}}{\Gamma _{L,e}+\Gamma
_{L,g}}$ is the injection efficiency into the ES. The filling rate, $F\simeq
\Gamma _{L,g}+\Gamma _{L,e}$, is so large (about 1 GHz in our experiment)
that e$^{-Ft}\sim 0$ in our time domain. The decay rate, 
\begin{equation}
D=W+\Gamma _{R,e}(1-A),  \label{EqW}
\end{equation}
contains information about $W$. The average number of tunneling electrons, $%
\langle n_{e}\rangle $, can be written as $\langle n_{e}\rangle \simeq \int
\Gamma _{R,e}\rho _{e}dt=A\Gamma _{R,e}(1-$e$^{-Dt})/D$, and can be compared
with the experiments described below.

In addition to the transient current, we can also measure the stable current
through the GS as a reference. The ES and GS become empty before the pulse,
and only the GS contributes to the stable current during the pulse [Fig.
1(f)]. Because there is no current-blocking for the stable current, the
average number of tunneling electrons, $\langle n_{g}\rangle $, should be $%
\langle n_{g}\rangle =\Gamma _{R,g}t$.

We apply a square-like pulse waveform through a dc-block (low frequency cut
off at 700 Hz) and a low-loss coaxial cable (about 3 m long with a loss of $<
$ 3 dB for dc - 10 GHz). There is no termination near the sample, and the
reflected pulse, as shown in Fig. 1(b), is more or less the actual pulse
waveform at the gate. The rise time (about 0.2 ns) of the pulse is faster
than $\Gamma _{L,g}^{-1}$ (typically about 1 ns), so the GS is not
immediately occupied. The pulse, whose length, $t_{p}$, is varied from 10 ns
to 10 $\mu $s, is repeated at a repetition time, $t_{r}$, and the average
current is measured.

Figure 2(b) shows one of the CB oscillations under pulse excitation. The
bias voltage ($V_{b}$ = 0.1 mV) is sufficiently large to saturate the
current, but is still small to enhance energy resolution. The application of
the pulse signal splits the single peak into two. Peak $g$ corresponds to
the stable current through the GS for the duration of the pulse ($t_{p}$),
and the large peak on the far right is also related to the current through
the GS but only for the duration of\ $t_{r}-t_{p}$. The extra peak $e_{1}$
is the transient current through the first ES, and it appears between the
two split peaks when the pulse height, $V_{p}$, is increased further. The
spacing between peaks $g$ and $e_{1}$ corresponds to the energy spacing
between the GS and the ES. Each peak broadens with increasing $V_{p}$.
Probably this is due to the weak ringing structure seen in the pulse
waveform or heating effects. The pulse signal also induces extra pumping
current, which is very slightly visible between two peaks in the trace at $B$
= 1.4 T in Fig. 2(b) [also visible as blue region in Fig. 3(c)].

We investigated the transient and stable currents as a function of the pulse
length, $t_{p}$. We can obtain the average numbers of tunneling electrons
per pulse, $\langle n_{g}\rangle $ for the stable current through the GS,
and $\langle n_{e}\rangle $ for the transient current through the ES from
the $It_{r}/e$ value for each peak current $I$. Figures 2(c) and (d) are the
results from different electron numbers in the dot. For both cases, $\langle
n_{g}\rangle $ depends linearly on $t_{p}$, indicating a continuous current
flow. Because tunneling rates are different, the characteristic time scale
is different. For the transient current, $\langle n_{e}\rangle $ shows
saturation behavior, which can be fitted well with the single exponential
function, $\sim (1-$e$^{-t/\tau })$. Data for a constant duty ratio ($%
t_{p}/t_{r}$ = 0.2) and for a fixed repetition time ($t_{r}$ = 200 ns)
coincide very well [Fig. 2(c)].

In principle, $W$ can be derived from $D=1/\tau $ using Eq. \ref{EqW}. We
can experimentally derive the tunneling rates, $\Gamma _{R,g}$ and $\Gamma
_{R,e}$ from the dc measurement, but $\Gamma _{L,e}/\Gamma _{L,g}=\Gamma
_{R,e}/\Gamma _{R,g}$ is assumed in order to obtain $A$. The $\Gamma
_{R,e}(1-A)$ is estimated to be about 20 MHz for $D=$ 22 MHz in Fig. 2(c),
and 0.8 MHz for $D=$ 0.5 MHz in Fig. 2(d). In both cases and also for most
of the ESs that appear in the transient current [see cross-hatch regions in
Fig. 3(d)], $W$ is too small to estimate with this technique, and we can
conclude the relaxation time is longer than, at least, 2 $\mu $sec for the
ES.

By contrast, if the relaxation is very fast ($W\gg \Gamma _{R,e}$ and $%
Wt_{p}\gg 1$), the transient current, $I=eA\Gamma _{R,e}/Wt_{r}$, becomes
very small. The upper bound of a measurable $W$ is 0.2 - 1 GHz under our
experimental conditions (current sensitivity of 50 fA and shortest $t_{p}$
of 10 ns). We find a small transient current for the specific ES $\epsilon $
[see Fig. 3(d)] at $B$ = 1.1 - 1.2 T, and estimate the relaxation time to be
about 3 ns. This is the only case where we were able to determine the actual
relaxation time. Other ESs missing in the transient current spectroscopy can
have a shorter relaxation time. Note that we cannot rule out other
experimental reasons for the absence of other ESs, namely that the injection
efficiency $A$ can be extremely small, or that the peak is greatly broadened
by the pulse.

The relaxation is either too fast or too slow to allow the relaxation time
to be estimated by our technique. We argue that the significantly different
relaxation times are the result of spin conservation during the relaxation.
If the ES and GS have the same spin, the electron in the ES quickly relaxes
to the GS as a result of electron-phonon interaction (typically about 1 ns). 
\cite{Bockelmann,Agam} By contrast, if they have different spins, the
relaxation time is limited by the spin-flip process. Spin coherence time in
bulk GaAs can exceed 100 ns.\cite{Awschalom} Most of the spin relaxation
mechanisms discussed for 3D or 2D electrons in GaAs are suppressed for 0D
states of QDs. We expect an extremely long spin relaxation time ($>$ 10 $\mu 
$s) due to a small admixture of different spin states.\cite{Khaetskii} This
is beyond our experimental limit (a few $\mu $s), and explains the
long-lived ESs we observed. In this case the long-lived ES should have
different spins from that of the GS. In our experiment, the Zeeman splitting
is small and unresolved. Thus the difference in the total spin is
responsible for the long energy relaxation time and for the appearance of
the transient current.

We can also confirm the existence of the long-lived ESs qualitatively from
the dc current,\cite{TOLK} when the outgoing tunneling rate is much smaller
than the incoming tunneling rate (simply making the bias voltage large and
positive in our experiment). An electron in the ES escapes to the right lead
either directly at rate $\Gamma _{R,e}$ or via the GS [see energy diagram of
inset in Fig. 3(b)]. If $W\gg \Gamma _{R,e}$, the electrons can always
escape from the GS. Thus the current is given by $e\Gamma _{R,g}$,
regardless of whether the ES is located in the transport window or not. If $%
W<\Gamma _{R,e}$ and $\Gamma _{R,e}\neq \Gamma _{R,g}$, the transport
through the ES can be distinguished from that through the GS. In this case,
another current step appears, for example, at $\mu _{L}=E_{e1}$ as indicated
in Fig. 2(a). It should be noted that the magnitude of a current step for
the ES is a complicated function of the tunneling rates,\cite{TFtobePB} and
a simple discussion based on the dc current can sometimes be unreliable.
Nevertheless, we believe that a qualitative analysis of the dc
characteristics is useful if the tunneling rates are adjusted properly to
distinguish the spin states. The $B$-dependence of the spectrum $dI/dV_{R}$
at $V_{b}$ = 1 mV [Fig. 3(b)] is almost the same as that found from
transient current spectroscopy [Fig. 3(c)], although the peaks in Fig. 3(c)
are broadened by the pulse irradiation.

We can speculate on the characteristics of the energy states from their $B$%
-dependence summarized in Fig. 3(d). At $B$ = 1.2 - 2 T, where edge states
are developing in the dot, some states ($\alpha $, $\beta $, $\gamma $, $%
\beta ^{\prime }$, $\gamma ^{\prime }$, and $\delta ^{\prime }$) are
energetically decreasing with increasing $B$. These states can be assigned
to the lowest Landau level, although they have different angular momentum
(slightly different slopes in the $B$-dependence).\cite{LPKexcitationAA}
Some of the ESs ($\beta $ and $\gamma ^{\prime }$) have an extremely long
relaxation time (Fig. 2(d) is taken for the state $\beta $ at $B$ = 1.35 T),
even though the GS and the ES are spatially overlapped.\cite{SingleDotVLVR}
It has been reported that the relaxation time is long if the two states are
spatially separated \cite{FujisawaScience} or when the two states are
attributed to different edge states,\cite{vanderVaart,Komiyama} but this is
not the case here.

Although we do not see clear spin-pairing in the GSs, specific ESs are
parallel to the GSs (e.g., the GS $\zeta $ and the ES $\xi $ at $B$ = 0.2 -
0.6 T). A parallel pair of the states is likely to have the same orbital
character but with different total spin $S$ due to exchange interactions. 
\cite{TaruchaAA} Actually the ES $\xi $ exhibits a transient current with a
relaxation time longer than 70 ns. The signal in the dc transport with small
outgoing tunneling rate [Fig. 3(b)] also indicates a long lifetime for the
ES $\xi $. This evidence supports the idea of spin conservation.

Conversely, the QD spin state can be investigated using transient current
spectroscopy based on the spin conservation rule. The transient current has
a long time constant for the first ES if it has a different total spin from
the GS, and also generally for the higher ES if it has a different total
spin from any lower-lying states. We see some long lived ESs cross with the
GS (e.g. the state $\xi $ becomes GS at $B\sim $ 0.9 T), indicating that the
total spin of the GS changes at this $B$. The simple spin-conservation
model, however, cannot explain the level crossing itself. The long-lived ESs
are sometimes terminated by a crossing with the GS or with another
(short-lived) ES. If the total spin does not change at the crossing, the
long-lived ESs should also remain, but this is not the case. The problem of
sensitivity must be studied further.

In summary, we investigated the energy relaxation time of ESs in QDs using a
pulse-excitation measurement. We found that the spin relaxation time is
longer than, at least, few $\mu $s, which is attractive for spin-related QD
applications.

We thank D. G. Austing, A. V. Khaetskii, L. P. Kouwenhoven, K. Muraki, M.
Stopa, and S. Tarucha for valuable discussions and help.

% now the references. delete or change fake bibitem. delete next three
%   lines and directly read in your .bbl file if you use bibtex.

\begin{figure}[tbp]
\epsfxsize=3.0in \epsfbox{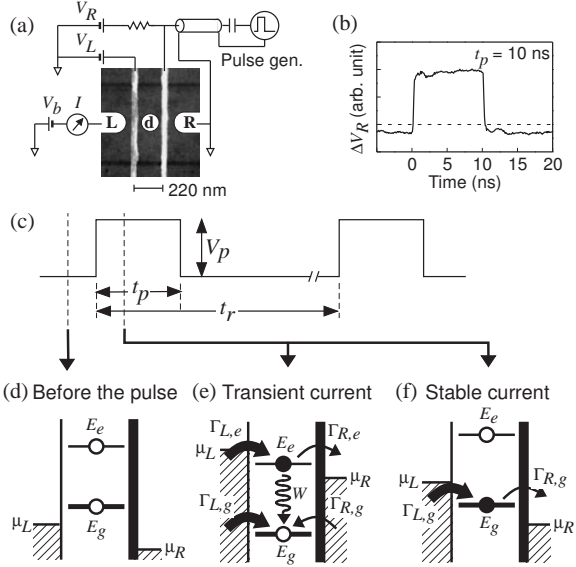}
\caption{(a) Schematic setup for the pulse measurements. All the
measurements are performed at a temperature of 150 mK in a magnetic field, $%
B$ = 0 - 2 T, perpendicular to the substrate. (b) Typical pulse waveform
obtained by time-domain reflectometry. (c) Schematic pulse train used in the
measurement. (d)-(f) Schematic energy diagrams. (d) Both the excited state
(ES) of energy $E_{e}$, and the ground state (GS) of energy $E_{g}$
become empty before the pulse. (e) The transient current through the ES
continues until the GS is occupied. (f) The stable current through the GS
persists during the pulse.}
\end{figure}

\begin{figure}[tbp]
\epsfxsize=3.0in \epsfbox{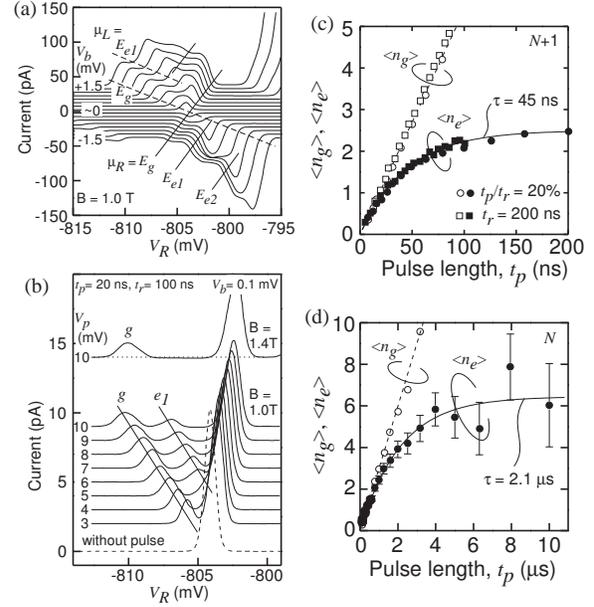}
\caption{(a) Single-electron tunneling current for various bias voltages, $%
V_{b}$. Each trace is shifted by 5 pA for clarity. (b) Pulse-excited current
for various pulse height, $V_{p}$. Each line
is shifted for clarity. At 1.0 T, the stable current $g$ through the GS (see Fig.
1(f)) and the transient current $e_{1}$ through the first ES (see Fig. 1(e))
are clearly seen. The upper trace measured at 1.4 T shows no transient
current. (c) and (d) The average number of tunneling electrons per pulse, $%
\langle n_{g}\rangle $ for a stable current through the GS (open symbols)
and $\langle n_{e}\rangle $ for a transient current through the ES (filled
symbols). The dashed and the solid lines are fitted to the data with a
linear and an exponential function, respectively. Tunneling rates for (d) 
are much smaller than that for (c). Conditions: (c) is at $B$
= 1.1 T for the $N+1$ electron state ($\Gamma _{R,g}$ = 26 MHz, $\Gamma
_{R,e}$ = 100 MHz, $\Gamma _{L}/\Gamma _{R}$ $\sim $ 15) and (d) is at $B$ =
1.35 T for the $N$ electron state ($\Gamma _{R,g}$ = 1.1 MHz, $\Gamma _{R,e}$
= 4.2 MHz, $\Gamma _{L}/\Gamma _{R}$ $\sim $ 300).}
\end{figure}

\begin{figure}[tbp]
\epsfxsize=3.2in \epsfbox{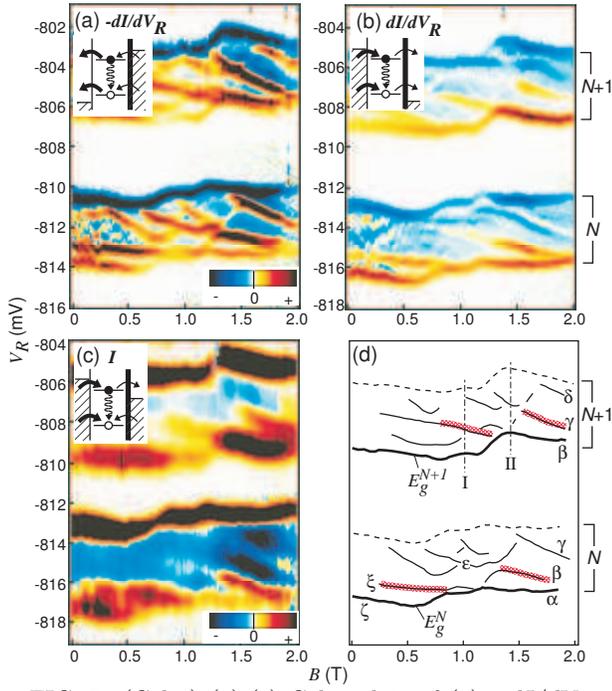}
\caption{(Color) (a)-(c) Color plots of (a) $-dI/dV_{R}$ at $V_{b}$ = -1 mV
(small incoming rate), (b) $dI/dV_{R}$ at $V_{b}$ = -1 mV (small outgoing
rate), and (c) pulse-excited current, $I$ at $V_{b}$ = 0.1 mV, $V_{p}$ $%
\simeq $ 10 mV and $t_{p}$/$t_{r}$ = 20 ns/100 ns. The upper and lower
stripes correspond respectively to the $N+1$ and $N $ electron states in the
dot. To obtain the best contrast in images (a)-(c), we multiplied the trace
at each $B$ by a smoothly varying function of $B$, and magnified the
intensity of the lower stripe by a factor of eight. (d) Traces of GS and ESs
extracted from (a)-(c). The thick lines, $E_{g}^{N}$ and $E_{g}^{N+1}$, represent the
GS energies. The thin lines represent their ESs. The dashed line is the upper
boundary of the excitation spectrum determined by $|V_{b}|$. The transient
current observed near the crosshatched regions exhibits a long relaxation
time. Vertical dot-dash lines, I and II, indicate the conditions used in Fig. 2(a)
and (b). }
\end{figure}


\begin{references}
\bibitem{SingleDotVLVR}  The excitation energy spectrum remains unaffected
when $V_{L}$ is changed by 15 mV, so all the states have a common set of
gate capacitances. This is also evidence that the GS and the ESs overlap
spatially.
\end{references}
\end{document}